\documentclass[10pt,conference]{IEEEtran}

\usepackage[utf8]{inputenc}
\usepackage{xspace}
\usepackage{hyperref}
\usepackage{graphicx}
\usepackage{xcolor}
\usepackage{enumerate}
\usepackage{balance}

\graphicspath{ {./images/} }

\newcommand{\toolname}{\textsc{Revizor}\xspace}

\newcommand{\ie}{\textit{i}.\textit{e}., }
\newcommand{\eg}{\textit{e}.\textit{g}., }

\makeatletter
\newcommand*{\rom}[1]{\expandafter\@slowromancap\romannumeral #1@}
\makeatother

\makeatletter
\newcommand{\linebreakand}{%
  \end{@IEEEauthorhalign}
  \hfill\mbox{}\par
  \mbox{}\hfill\begin{@IEEEauthorhalign}
}
\makeatother

\title{\toolname: A Data-Driven Approach to Automate Frequent Code Changes Based on Graph Matching}

\author{
 \IEEEauthorblockN{Oleg Smirnov,\IEEEauthorrefmark{1}\IEEEauthorrefmark{2} Artyom Lobanov,\IEEEauthorrefmark{1}\IEEEauthorrefmark{4} Yaroslav Golubev,\IEEEauthorrefmark{1} Elena Tikhomirova,\IEEEauthorrefmark{1} Timofey Bryksin\IEEEauthorrefmark{1}\IEEEauthorrefmark{2}\IEEEauthorrefmark{4}}
    \IEEEauthorblockA{\IEEEauthorrefmark{1}\textit{JetBrains Research},  \IEEEauthorrefmark{2}\textit{Saint Petersburg State University},
    \IEEEauthorrefmark{4}\textit{Higher School of Economics}}
    \IEEEauthorblockA{\{oleg.smirnov, artem.lobanov, yaroslav.golubev, elena.tikhomirova, timofey.bryksin\}@jetbrains.com}
}

\begin{document}

\maketitle

\begin{abstract}

Many code changes that developers make in their projects are repeated and constitute recurrent change patterns. It is of interest to collect such patterns from the version history of open-source repositories and suggest the most useful of them as quick fixes. In this paper, we present \toolname---a tool aimed to build custom plugins for PyCharm, a popular Python IDE. A \toolname-based plugin can take change patterns and highlight potential places for their application in the developer's code editor. If the developer accepts the quick fix, the plugin automatically performs the edit. Our approach uses a graph-based representation of code changes, which allows it to support complex distributed code patterns. Experienced developers have also rated the usability and the performance of such \toolname-based plugin positively.

The source code of the tool and test plugin prototype are available on GitHub: \url{https://github.com/JetBrains-Research/revizor}. A demonstration video with a short tool description can be found on YouTube: \url{https://youtu.be/5eLs14nco7E}.

\end{abstract}

\section{Introduction}

In recent decades, Integrated Development Environments (IDEs) have been evolving to provide tools that help developers write and edit code ever more efficiently~\cite{zayour2013much}. 
Writing code involves making a lot of incremental changes and small fixes. Some of such scrupulous high-concentration tasks can be automated to boost developers' performance. 
To this end, IDEs incorporate \textit{static analysis tools} that find and highlight known problems in code, as well as suggest certain \textit{quick fixes} as possible solutions to these problems in real time.
A lot of works focus on automated program repair (APR)~\cite{monperrus2020living}, which includes localizing bugs and vulnerabilities in code and fixing them. 
We aim to show that approaches which are similar to APR can be applied not only to bug fixes, but to almost any code change that is of interest to developers---from stylistic code enhancement to the migration of APIs to a different language version.

A good place to search for possible improvements of code is the history of it being changed. 
Prior research suggests that code changes are repetitive: from time to time different developers not only face the same kind of problems, but try to solve them in a similar way~\cite{nguyen2013study, nguyen2019graph}.
Such code changes constitute recurrent change patterns that can be mined from the histories of existing software projects.

Semantic change patterns often differ in structure significantly. While some of them are one-liners or even touch only a single statement~\cite{karampatsis2020often}, others can be quite complex and \textit{distributed}, \ie involving isolated tokens from different lines of code or even different scopes connected by data or control dependencies (as in Figure~\ref{fig:example}).
From that point of view, code analysis approaches can benefit greatly from utilizing the fact that code is more than just plain text and work with graph-based code representations that make it possible to keep track and make use of the code's structure and make decisions about its semantics~\cite{nguyen2019graph, allamanis2017learning}.

In this paper, we investigate whether it might be possible to use graph-based representations of recurrent change patterns for code improvement suggestions within an IDE. 
For that purpose, we have developed \toolname---a tool that allows to build code enhancement plugins for PyCharm~\cite{PyCharm}, a popular IDE for Python developed by JetBrains. We also propose our prototype plugin built with nine pre-approved code change patterns. 
For any given change pattern, the plugin uses a fine-grained program dependence graph (\textit{fgPDG} introduced by Nguyen et al.~\cite{nguyen2019graph} for mining patterns in Java) of the code fragment before the change to localize its occurrences in the user's code via building subgraph isomorphisms.
If such an occurrence is found, the plugin highlights the code fragment for the user, indicating that a quick fix can be applied. 
If the user chooses to apply the fix, the plugin can do it by performing a sequence of edit actions, which are generated for each pattern from the versions of code before and after the change.

To evaluate the proposed plugin, we conducted a survey of nine experienced developers. 
The participants positively rated the plugin's usability and performance and mostly approved the idea of using recurrent code changes to suggest relevant quick fixes in the IDE.
\section{Background}

In this section, we describe the approaches that we adopted to build our pipeline. 
In Sections~\ref{background-localization} and \ref{background-application}, we characterize the solutions that we considered for the plugin to localize patterns in source code and to apply the respective changes (see the two final stages in Figure~\ref{fig:pipeline}), and we also contrast the adopted solutions with similar ones. 
In Section~\ref{background-collection}, we relate a prospective graph-based approach to code change patterns mining. 
Integrating it into a code enhancement pipeline is unprecedented and seems promising. 

\subsection{Code Pattern Localization}
\label{background-localization}

There exists a number of relevant approaches that aim to mine fix patterns and use them to localize potential code flaws.

Meng~et~al. presented an approach called \textsc{LASE}~\cite{meng2013lase}, where the authors initially built a \textit{context-aware edit script} from two or more code change examples and then identified appropriate locations for code transformation with a \textit{generalized tree-based edit context}.
To localize such a context in the AST of target code, the authors used the Maximum Common Embedded Subtree Extraction algorithm.
A similar approach to localization with a tree-matching algorithm was also used by Bader~et~al. in their tool called \textsc{Getafix}~\cite{bader2019getafix}, in which bug fix patterns mined in Java are automated.
However, these approaches cannot be used to localize \textit{distributed} code patterns, which may include many subtrees of an AST.

The authors of \textsc{DevReplay}~\cite{ueda2020devreplay} collected code change patterns via AST comparison, converted the source code of the patterns into regular expressions, and then tried to match any of them with the user's code to localize possible problems in it.
Such method is able to handle multi-line patterns, but still cannot automatically manage control or data flow dependencies between elements of the pattern as graph-based approaches do. 

\subsection{Edit Template Application}
\label{background-application}

After a pattern is localized in code and the developer confirms applying the fix, the exposed code fragment is changed in accordance with the appropriate edit script.
This is usually done via AST transformations similarly to how it was described by Meng et al.~\cite{meng2013lase, meng2011systematic}. 
For that purpose, the authors used edit actions of four types---\textit{insert}, \textit{delete}, \textit{update} and \textit{move}---generated by a modified version of \textsc{ChangeDistiller}~\cite{fluri2007change}, a source code differencing tool for extracting fine-grained edit scripts from two versions of an AST: before and after the change.

Bader et al.~\cite{bader2019getafix} used a similar tool called \textsc{GumTree}, and, according to Falleri et al.~\cite{falleri2014fine}, \textsc{GumTree} represents edit actions in a more accurate and concise way compared to other source code differencing tools.

Nowadays, researchers also widely exploit the ideas of Neural Machine Translation (NMT) that view the task of applying changes as a translation problem from a defective code fragment into a correct one~\cite{tufano2019empirical, lutellier2019encore}.
These approaches are hardly interpretable and require collecting a large number of similar patches to train the model, which is a difficult task.
On the contrary, heuristic approaches like the ones that employ AST edit actions need far less input and computational resources to perform.

\subsection{Collection of Recurrent Code Changes Using Graphs}
\label{background-collection}

To keep track of semantic features, as well as data and control dependencies between the elements of source code, more complex data structures such as graphs can be used.
Nguyen et al. proposed an approach called \textsc{CPatMiner}~\cite{nguyen2019graph} for mining graph-based change patterns in Java code. 
The representation of code that they used is called \textit{Fine-Grained Program Dependence Graph (fgPDG)} and is based on the AST of the source code. Such a graph includes three types of nodes: data nodes (for variables, literals, etc.), operation nodes (for arithmetic expressions, assignments, function calls, etc.), and control nodes (for control statements like \texttt{if}, \texttt{for}, \texttt{while}, etc.). 
These nodes are linked with additional data and control dependency edges. 

To represent code changes, Nguyen~et~al. introduced the concept of a \textit{change graph}. 
A change graph is built using two fgPDGs of the code before and after a given change; corresponding unchanged graph nodes are connected with mapping edges. 
The authors also suggested a way to use this data structure to build a pattern-mining algorithm. The main idea behind it is to recursively extend each already mined change graph to the most frequently encountered adjacent vertex and then match isomorphic graphs using a hash-based heuristic~\cite{nguyen2009accurate} to put them into one particular pattern.

In our prior work~\cite{golubev2021changes}, we re-implemented this approach for Python, collected and analyzed fgPDG-based code change patterns from 120 popular GitHub repositories.
This allowed us to collect recurrent in-the-wild code changes: code enhancements, bug fixes, refactorings, etc. 
In this work, our goal was to implement quick-fixing actions so that these graph-based changes could be applied automatically in the developer's code in the IDE.

\begin{figure*}[ht]
    \centering
    \includegraphics[width=\textwidth]{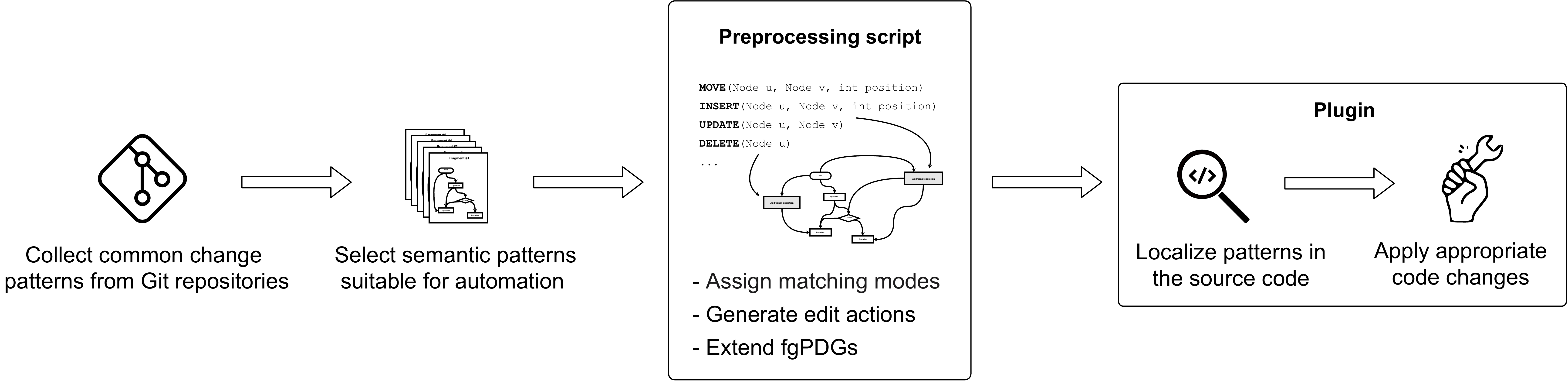}
    \caption{An overview of the proposed approach.}
    \label{fig:pipeline}
     \vspace{-5mm}
\end{figure*}
\section{Implementation}
\label{sec:implementation}

\subsection{Pipeline Overview}
We have implemented an approach to enhancing IDEs with valuable up-to-date code improvement suggestions in a data-driven way. 
Our contribution is \toolname, a tool that allows to create custom plugins with pre-approved quick fixes for Python code.  
We also built a prototype plugin for PyCharm, a popular Python IDE based on the IntelliJ Platform.\footnote{\url{https://plugins.jetbrains.com/docs/intellij/intellij-platform.html}} 
The plugin is created using code change patterns and respective code samples mined with PythonChangeMiner~\cite{golubev2021changes}, which analyzes the graph-based representation of code changes in Git repositories and detects change patterns without any prior specification of what is worth changing in Python code, thus sparing the necessity to devise and manually write code enhancement rules. 
The proposed approach allows plugins to locate \textit{distributed} code patterns, \ie the patterns involving isolated tokens that are located on multiple lines of code and connected by some data or control dependencies.
Using fgPDGs enables greater matching flexibility and structural awareness compared to general regex patterns.

The full pipeline behind \toolname is shown in Figure~\ref{fig:pipeline}.
The steps to build the plugin are as follows:
\begin{enumerate}
    \item \textit{Collecting}: Collect graph-based patterns of code changes in Python and choose the ones that should be automated.
    \item \textit{Preprocessing}: Build the \toolname plugin using any type of sources from step 1. If in future any new change patterns are added, re-build the plugin.
\end{enumerate}
The steps are described in more detail in the next subsections. For more information on how to build a plugin with \toolname, see its README on GitHub~\cite{revizor}.

After the plugin is installed in the IDE, it tracks developer's actions when a \texttt{.py} file is opened or changed in the code editor.
Namely, the plugin:
\begin{enumerate}[i]
    \item Builds an fgPDG for each function in the developer's code using its PSI tree (an enriched form of a concrete syntax tree used in the IntelliJ Platform).\footnote{\url{https://plugins.jetbrains.com/docs/intellij/psi.html}} 
    It is performed on-the-fly using IntelliJ's mechanism called \textit{code inspections}.\footnote{\url{https://plugins.jetbrains.com/docs/intellij/code-inspections.html}}
    \item Checks such graphs for possible subgraph isomorphisms with the \textit{before} version of each available change pattern stored in the plugin's resources.
    \item Highlights the corresponding code tokens in the editor if such an isomorphism is detected and suggests the respective improvement.
    \item Performs the improvements confirmed by the developer using a sequence of \textit{edit actions} extracted with \textsc{GumTree} during the preprocessing step.
\end{enumerate}

\subsection{Change Pattern Collecting}
\begin{figure}[t]
    \centering
    \includegraphics[width=\linewidth]{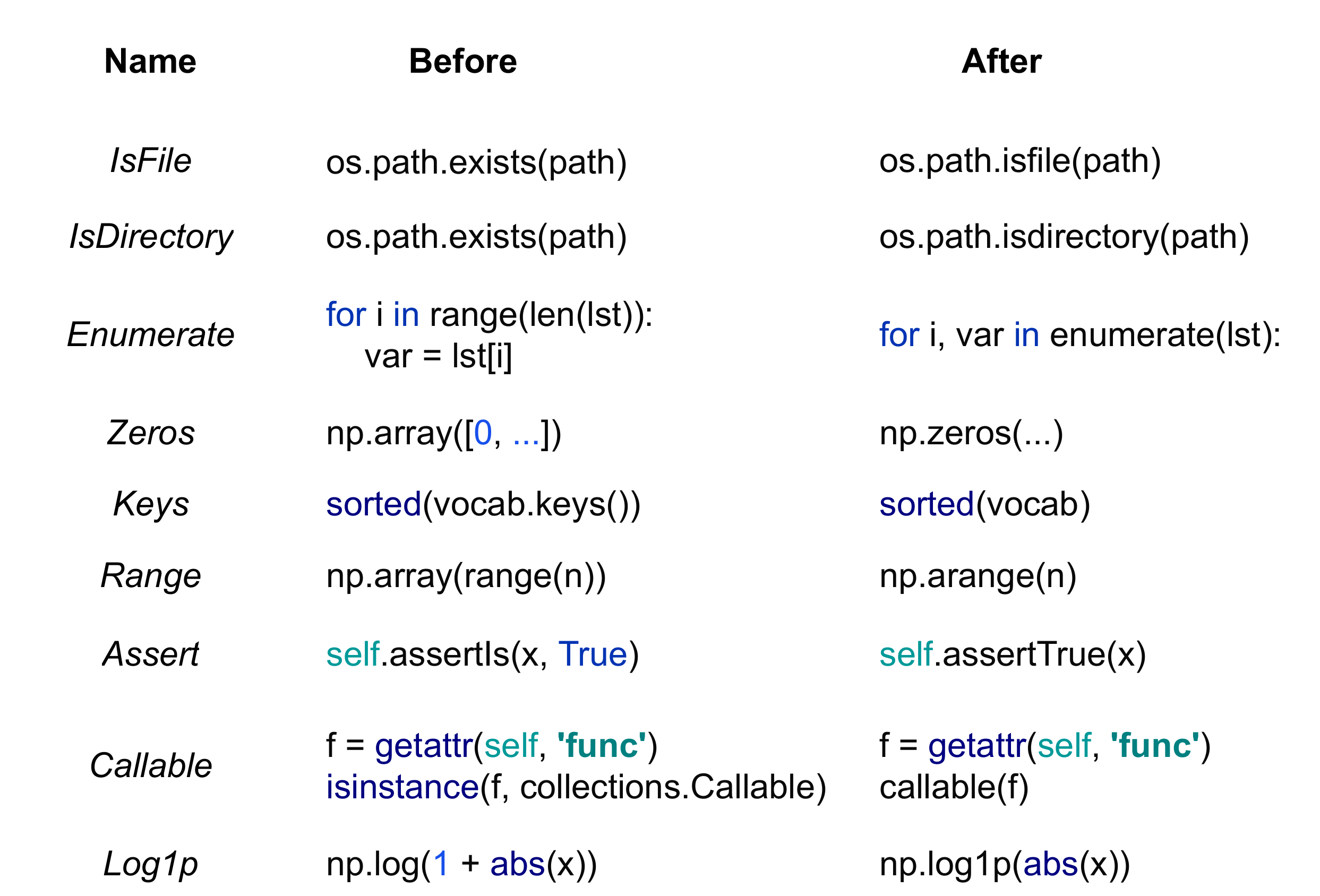}
    \caption{The examples of the chosen nine change patterns for \toolname-based plugin prototype evaluation.}
    \vspace*{1.2mm}
    \label{tab:patterns}
    \vspace{-4mm}
\end{figure}

\begin{figure}[t]
    \centering
    \includegraphics[width=\linewidth]{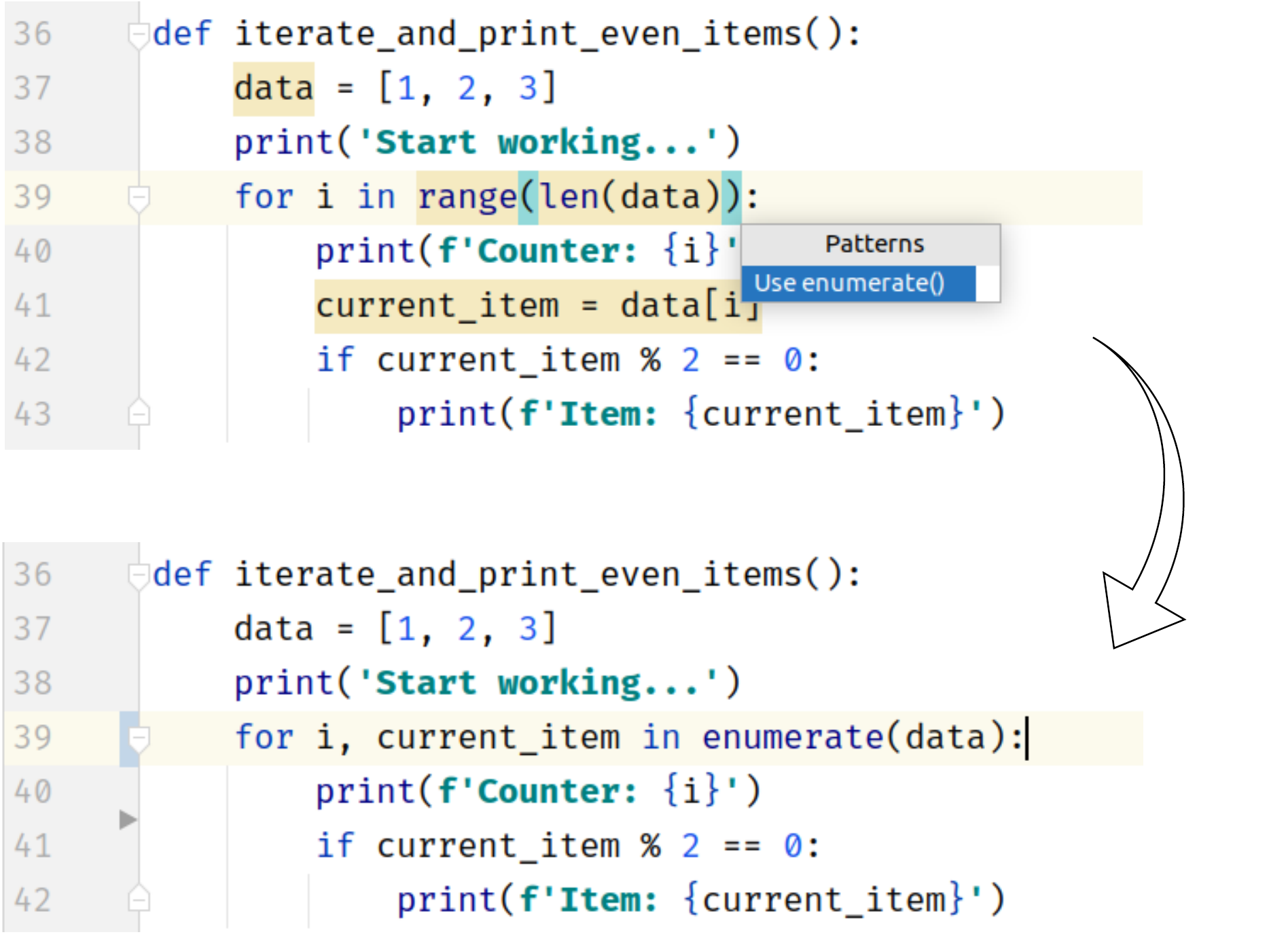}
    \caption{An example of applying a graph-based change pattern in our plugin: replacing the \texttt{range} function call in the \texttt{for} loop condition with \texttt{enumerate}. The highlighted code tokens are placed in different lines of code and include all the vertices of the detected fgPDG with respect to the data flow dependencies, \eg \texttt{data} declaration in line 37.}
     \vspace{1.2mm}
    \label{fig:example}
     \vspace{-5mm}
\end{figure}

In our prior work~\cite{golubev2021changes}, we had gathered a dataset of 120 popular GitHub repositories based on their domain, length of the commit history, age of the project, and number of contributors. 
Finally, we had discovered a total of 7,481 code change patterns. 
To understand their semantics better, we manually evaluated and categorized 803 patterns that appeared in at least two projects.
From this pool, we selected nine patterns presented in Table~\ref{tab:patterns} for the evaluation of our test plugin's prototype according to the following criteria: the patterns had different structure and semantics and constituted good examples of what software engineers we consulted thought worth automating.
The selected changes are related to developers' best practices, which evolve rapidly with any language and are rarely documented promptly. 
Also, fixes for such changes as \textit{Enumerate} are not easy to implement because the pattern could be distributed. \toolname-based plugins not only detect such unique change patterns but fix them as shown in Figure~\ref{fig:example}.
Finding change patterns can be automated to a large extent and it is a promising direction of future work.

\subsection{Pattern preprocessing}
\label{subsection:processing}
Before the mined change patterns can be used in the plugin, they should be preprocessed with our tool.
The process is fully automated except for the specification of tooltip annotations for each pattern that will appear in the IDE.

\subsubsection{Assignment of matching modes}
During preprocessing of the supplied graphs, \toolname automatically specifies how \textit{data} vertices from the pattern should be matched with the ones from developer's code when a \toolname-based plugin looks for ``familiar'' patterns in code. Such rules are called \textit{matching modes} and take into account the vertices' labels, positions, and neighbours in the fgPDG.

\textit{Examples}. Some \textit{user-defined} variable names that refer to the same data element (\eg a list may be named \texttt{lst}, \texttt{data} or \texttt{items}) do not need to be exactly matched during subgraph isomorphism search, and therefore we assigned such vertices with the \texttt{match$\_$any$\_$label} matching mode.
Other variable names \textit{should} be considered as having a partial match with a common suffix, \eg \texttt{dict.keys} and \texttt{vocab.keys} (the \texttt{match$\_$longest$\_$common$\_$suffix} mode). Some should always be matched exactly as they are built-in or external library Python functions and attributes, \eg \texttt{collections.Callable} or \texttt{np.log} (the \texttt{match$\_$original$\_$labels} mode).

\subsubsection{Generating edit actions}
We use \textsc{GumTree}~\cite{falleri2014fine} to extract sequences of edit actions from PSI trees of the before and after code fragments related to the change (they are stored by the miner together with the respective graphs). Edit actions keep references to the corresponding vertices of the PSI tree before the change, making it possible to apply these actions to code merely one by one.

While \textsc{GumTree} extracts \textit{all} edits from such code changes (most of them potentially unrelated to our pattern), we need to get only \textit{necessary} actions. We do so by calculating a Longest Common Edit Operation Subsequence with generalized identifiers by iteratively comparing the extracted edit actions sequences pairwise.
A similar process was described in detail in the work about \textsc{LASE}~\cite{meng2013lase}, but instead of keeping an edit context, we use the isomorphic mappings between fgPDGs for all encountered fragments of the pattern.

\subsubsection{Extending fgPDGs}
When an appropriate subsequence of edit actions is extracted and saved, we extend the originally mined fgPDG of the pattern with additional vertices that appeared in these edits.
This is done because some of them may contain PSI nodes that did not even exist in the original fgPDG of the pattern, such as parents of the \textit{moved} or \textit{inserted} vertices in the PSI tree.

Finally, the preprocessing script automatically saves the assigned matching modes and edit actions, as well as the extended graph and the manually provided description of each pattern. 
After loading all of them as resources, the plugin is ready to go.

\subsection{Possible Application}

A \toolname-based plugin may meet the following purposes of a code standardization linter:
\begin{itemize}
    \item Self-education (individual level): Use our prototype with a selection of promising code changes from popular GitHub repositories.
    \item Enforcement of style guidelines (team/company level): Build a plugin around mined or manually created patterns of interest.
    \item Introduction of fresh high-quality inspirations from other developers (individual/team/company/education level): Build a plugin around mined and sifted patterns from relevant code repositories.
\end{itemize}

\section{Evaluation}

As a preliminary study, we asked nine developers to install our plugin in their PyCharm IDE and test it on an example project~\cite{exampleproject}, which contained manually selected code snippets from several Python projects where the chosen patterns were encountered during mining.
The participants were requested to find and perform all the suggested code changes, considering the usability of the tool.
All the developers had from two to five years of professional experience and confirmed that they often used intention actions in the IntelliJ-based IDEs to improve their code quality. 
The survey participants also agreed that the idea of using the most common code changes mined from GitHub as quick fixes in the IDE looked potentially useful.

We asked them to rate different aspects of the plugin's performance with one of the four responses: \textit{very dissatisfied} (1 point), \textit{not really satisfied} (2 points), \textit{rather satisfied} (3 points) and \textit{very satisfied} (4 points). 
The average score we received regarding the correctness of the plugin's edit operations was 3.66 out of 4, and the overall usability was rated at 3.77 out of 4. 
Developers also highly evaluated the performance of \toolname in terms of not affecting the overall IDE performance (3.88 out of 4). 
The lowest rated feature of the plugin was the pattern's visualization part (3.33 out of 4), as it turned out that the current approach to highlighting complex distributed patterns sometimes could be confusing for the developers.

\section{Conclusion and Future Work}
In this paper, we presented an extendable approach to automated code enhancement.
We proposed a data-driven tool called \toolname for building static code analysis plugins that use subgraph isomorphism mappings for pattern localization and \textsc{GumTree} edit actions to automate changes. 
The tool uses frequent Python change patterns mined from GitHub repositories.
We also created a test prototype of the plugin for a popular Python IDE called PyCharm and evaluated it on several experienced developers who approved its usability.

We received a lot of feedback about improving the UI/UX of the plugin, \ie how it treats the patterns, including an idea to highlight distributed patterns in a more intelligent way: first highlight the key token and when the user clicks it, highlight the other parts of the pattern. Also, we received feature requests such as to apply the selected change across the whole project or to suppress the suggestions for particular scopes of code. We aim to implement these features in future.

Overall, the described data-driven approach is potentially extendable to any other programming languages, but in order to capture any data or control dependencies in the code, the extended approach should be tightly dependent on the language grammar.
We also plan to support the unified fgPDG representations in our tool for other languages using the PSI. 

\balance

\bibliographystyle{ieeetran}
\bibliography{cites}

\end{document}